\documentclass[a4paper]{JHEP3}

\usepackage{bbm}
\usepackage{bm}
\usepackage{graphicx}
\usepackage{epsfig}
\usepackage{subfigure}
\usepackage{latexsym}
\usepackage{amsmath}
\usepackage{amsfonts}
\usepackage{amssymb}
\usepackage[utf8x]{inputenc}

\newcommand{\Tr}{\textrm{Tr}\,}

\newcommand{\csw}{c_{\rm{sw}}}

\newcommand{\fig}[1]{fig.~\ref{#1}}

\newcommand{\Dslash}{\ensuremath \raisebox{0.025cm}{\slash}\hspace{-0.28cm} D}
\newcommand{\be}{\begin{equation}}
\newcommand{\ee}{\end{equation}}
\newcommand{\bea}{\begin{eqnarray}} % only untightened
\newcommand{\eea}{\end{eqnarray}}

\newcommand{\bmp}{\noindent\begin{minipage}{16cm}}
\newcommand{\emp}{\end{minipage}\vskip 7mm} % 7mm untightened
\def\lsim{\mathrel{\raise.3ex\hbox{$<$\kern-.75em\lower1ex\hbox{$\sim$}}}}
\def\gsim{\mathrel{\raise.3ex\hbox{$>$\kern-.75em\lower1ex\hbox{$\sim$}}}}

\newcommand{\intron}[1]{}%{#1}
\title{Nonperturbative improvement of SU(2) lattice gauge theory 
  with adjoint or fundamental flavors}
\author{Tuomas Karavirta\footnote{tuomas.karavirta@jyu.fi}\\
Department of Physics, P.O.Box 35 (YFL), 
        \\ FI-40014 University of Jyv\"askyl\"a, Finland, 
        \\ and 
  	    \\ Helsinki Institute of Physics, P.O.~Box 64, 
  	    \\ FI-00014 University of Helsinki, Finland.}

\author{Anne Mykkanen\footnote{anne-mari.mykkanen@helsinki.fi}\\
Department of Physics and Helsinki Institute of Physics,\\
P.O.Box 64, FI-00014 University of Helsinki, Finland}
\author{Jarno Rantaharju\footnote{jarno.rantaharju@helsinki.fi}\\
 Department of Physics and Helsinki Institute of Physics,\\
 P.O.Box 64, FI-00014 University of Helsinki, Finland}
\author{Kari Rummukainen\footnote{kari.rummukainen@helsinki.fi}\\
 Department of Physics and Helsinki Institute of Physics,\\
 P.O.Box 64, FI-00014 University of Helsinki, Finland}
\author{Kimmo Tuominen\footnote{kimmo.tuominen@jyu.fi}\\
Department of Physics, P.O.Box 35 (YFL), 
        \\ FI-40014 University of Jyv\"askyl\"a, Finland, 
        \\ and 
  	    \\ Helsinki Institute of Physics, P.O.~Box 64, 
  	    \\ FI-00014 University of Helsinki, Finland.}
\abstract {%
  SU(2) gauge theory with two fermions transforming
  under the adjoint representation may appear conformal
  or almost conformal in the infrared, and is one of the
  candidate theories for building models for technicolor.
  Early lattice Monte Carlo studies of this model have
  used unimproved Wilson fermion formulation, which can
  be expected to have large lattice cutoff effects.  
  In this paper we present the calculation of the $O(a)$ improved
  lattice Wilson-clover action of the theory.  The Sheikholeslami-Wohlert
  coefficient has been determined non-perturbatively, and
  various boundary improvement terms, needed for the Schr\"odinger
  functional formalism, have been calculated in perturbation
  theory.  For comparison, we have also determined the improvement
  coefficients for SU(2) gauge theory with two fundamental
  representation fermions.
  The calculation paves way for more accurate lattice 
  Monte Carlo analyses of the theory in the future.
}

\keywords{Lattice field theory, Conformal field theory}
\begin{document}

%%%%%%%%%%%%%%%%%%%%%%%%%%%%%%%%%%%%%%%%%%%%%%%%%%%%%%%%%%%%%%%%%%%%%%%%%%%%%

\section{Introduction}

Quantum field theories with nontrivial infrared fixed points of the $\beta$-
function have recently been studied due to their applications in beyond 
Standard Model model building. In these theories the coupling runs when 
probed at very short distances, but becomes a constant over some energy range 
in the infrared and the theory appears conformal. One of the phenomenological 
connections is the unparticle \cite{Georgi:2007ek,Georgi:2007si,Sannino:2008nv}, 
i.e. the possibility of a fully conformal sector coupled only weakly to the
Standard Model through effective operators at low energies. 
Another phenomenological motivation to study theories which either
feature an infrared fixed point or are, in theory space, close to
one which does, originates from technicolor (TC) and the associated
extended technicolor (ETC) models. These models were devised 
to explain the mass patterns of the Standard Model gauge
bosons and fundamental fermions without the need to introduce a
fundamental scalar particle \cite{TC,Eichten:1979ah,Hill:2002ap,Sannino:2008ha}. 

Early TC models, based on a technicolor sector straightforwardly extrapolated 
from a QCD-like strongly interacting theory, lead to too large
flavor changing neutral currents due to the extended
technicolor interactions.The problems of these simple TC models are 
solved in so called walking technicolor theories 
\cite{Holdom:1981rm,Yamawaki:1985zg,Appelquist:an,Appelquist:1998rb}. 
These theories are quasi-conformal, i.e. the evolution of the coupling 
constant is, over a wide range of energy, governed by an attractive
quasi-stable infrared fixed point at strong coupling.

To build walking TC models one needs to tune the gauge and matter 
degrees of freedom so that the desired quasi-conformality arises. 
To achieve this in SU($N$) gauge theory with fermions in the 
fundamental representation several ${\cal{O}}(10)$ Dirac flavors 
are required. These contribute to the precision parameter $S$, which
becomes too large to be compatible with the current observations. To 
obtain enough screening, as required for quasi-conformality, but with 
smaller number of flavors, one considers fermions in higher representations. 
It has been suggested \cite{Sannino:2004qp} that an ideal candidate for 
minimal walking technicolor theory would be the one with just two 
(techni)quark flavors in the two-index symmetric representation of SU(2) or SU(3).  

Reliable quantitative studies of the models, especially
evaluating the $\beta$-functions, require lattice Monte Carlo 
simulations.
There are several recent studies 
of both SU(2)
\cite{Catterall:2007yx,Hietanen:2008mr,
  DelDebbio:2008zf,Catterall:2008qk,Hietanen:2009az,
  Bursa:2009we,DelDebbio:2009fd,DelDebbio:2010hx,
  DelDebbio:2010hu,Kerrane:2010xq}
and SU(3) 
\cite{Shamir:2008pb,DeGrand:2008kx,
  DeGrand:2010na,Fodor:2009ar,Kogut:2010cz}
gauge theories with two-index symmetric representation
fermions.  For related studies in QCD-like theories
with fundamental representation fermions see
\cite{Damgaard:1997ut,Appelquist:2007hu,Appelquist:2009ty,
  Fodor:2009wk,Deuzeman:2008sc,Deuzeman:2009mh,Itou:2010we,
  Jin:2010vm,Hayakawa:2010yn,Hasenfratz:2010fi,Hasenfratz:2009ea,
  Bursa:2010xr,Ohki:2010sr}.

In this paper we consider the case of SU(2) gauge fields with two
fermions in the two-index symmetric representation, which, for
SU(2), is equivalent to the adjoint representation. So far the
lattice studies of this theory have been performed
using unimproved Wilson fermion action and are hence subject to 
large ${\cal{O}}(a)$ lattice artifacts. 
In this paper we present the computation
of ${\cal{O}}(a)$-improvement.  This is a generalisation
of the program used earlier to compute the improved
action for two fundamental representation fermions in SU(3)
gauge theory \cite{Luscher:1992zx,Sint:1995rb,
  Sint:1995ch,Luscher:1996vw,Luscher:1996ug,Jansen:2008vs}.
The early results of this calculation have been presented
in refs.~\cite{Karavirta:2010ef,Mykkanen:2010ym}.

The Wilson fermion action can be improved for on-shell quantities by 
adding the well-known clover term.  We tune the coefficient of the
clover term (Sheikholeslami-Wohlert coefficient \cite{Sheikholeslami:1985ij})
non-perturbatively, using the Schr\"odinger functional method.
For the measurement of the coupling constant we also need the
improvement coefficients of certain boundary terms.  This computation
is done using perturbative analysis.  
For comparison, we also calculate the improvement for SU(2)
gauge theory with two flavors of fundamental representation
fermions.\footnote{%
  Non-perturbative improvement of the clover term 
  has been recently published for SU(3) gauge field
  theory with 2-index symmetric (sextet) fermions,
  using the HYP-smeared link clover action
  \cite{Shamir:2010cq}.}

The paper is structured so that in section~\ref{model} we first recall the
basics of the model as well as of the lattice formulation we use.
In section \ref{perturbative} we 
present our perturbative results for the boundary terms 
and nonperturbative results for the improvement
coefficients are presented in section \ref{cswdetermination}. 
In section \ref{outlook} we conclude and outline the
directions of our future work.

\section{Lattice formulation: the model and ${\mathcal{O}}(a)$
  improvement}
\label{model}

%\subsection{Preliminaries}

We study SU(2) gauge theory with two different matter contents: two
mass-degenerate flavors of Dirac fermions either in the adjoint or in
the fundamental representation.  The continuum theory in Euclidean
spacetime is defined by
\begin{equation}
  {\mathcal{L}}= 
  \frac{1}{2}\Tr F_{\mu\nu}F_{\mu\nu} + 
  \sum_{\alpha}\bar\psi_\alpha(i \Dslash + m)\psi_\alpha
  \label{eq:continuumL}
\end{equation}
where $F_{\mu\nu}$ is the usual SU(2) field strength, and the
gauge covariant derivative is
\begin{equation}
  D_\mu \psi =
  \left(\partial_\mu - ig A^a_\mu T^{a}\right) \psi
\end{equation}
where $a=1,2,3$ and the generators $T^a$ are taken either in the
fundamental ($T^a = \sigma^a/2$) or in the adjoint representation
($[T^a]^{bc} = -i\epsilon^{abc}$).  The summation in
Eq. (\ref{eq:continuumL}) is over $\alpha = {\textrm{u}},{\textrm{d}}$.

Our main goal in this work is to establish nonperturbative
${\mathcal{O}}(a)$ improved lattice implementation of these
theories.  While the improvement has been discussed in detail in existing
literature for SU(3) gauge field with fundamental fermions, the
studies of adjoint flavors require some alterations. Hence we find it
necessary and useful to repeat essential parts of the analysis in
detail here.

First recall the usual ${\mathcal{O}}(a)$ improvement obtained by
Sheikholeslami and Wohlert \cite{Sheikholeslami:1985ij}. The lattice
action, split to the gauge and fermionic parts $S_G$ and $S_F$, is
\begin{equation}
  S_{0}=S_G + S_F.
  \label{eq:aktion}
\end{equation}
Here we use the standard Wilson plaquette gauge action
\begin{equation}
 % S_G =  \beta_L \sum_p{\textrm{tr}}(1-U(p)).
  S_G = \beta_L \sum_{x;\mu<\nu} 
  \left (1 - \frac12 \Tr P_{x;\mu\nu}\right)
%  =\frac{\beta_L}{4}a^4\sum_{x} F_{\mu\nu}(x)F^{\mu\nu}(x)
%  +{\cal{O}}(a^6),
\end{equation}
where $\beta_L = 4/g_0^2$ and 
the plaquette is written in terms of the
SU(2) fundamental representation link matrices $U_\mu(x)$, which
act as parallel transporters between sites $x$ and $x + a\hat\mu$:
\begin{equation}
  P_{x;\mu\nu} = U_\mu(x) U_\nu(x+a\hat\mu) 
  U^\dagger_\mu(x+a\hat\nu) U^\dagger_\nu(x).
\end{equation}
The Wilson fermion action, $S_F$, for $N_f$ (degenerate) Dirac
fermions in the fundamental or adjoint representation of the gauge
group is
\begin{equation}
  S_{\textrm{F}}
  = a^4\sum_\alpha \sum_x 
  \bar{\psi}_\alpha(x)(iD + m_{q,0}{\mathbbm{1}})
  \psi_\alpha(x),
  % \sum_{N_f} 
  %\sum_{x,y} \bar\psi_{f,x} M_{xy} \psi_{f,y} 
    \label{sf}
\end{equation}
where the usual Wilson-Dirac operator is
\begin{equation}
  D=\frac{1}{2}(\gamma_\mu(\nabla_\mu^\ast+\nabla_\mu)
  - a \nabla^\ast_\mu\nabla_\mu),
\end{equation}
involving the gauge covariant lattice derivatives 
$\nabla_\mu$ and $\nabla_\mu^\ast$ defined as
\begin{eqnarray}
  \nabla_\mu\psi(x) &=& \frac{1}{a}[ \widetilde{U}_\mu(x) \psi(x+a\hat{\mu})
  - \psi(x)], \\
  \nabla^\ast_\mu\psi(x) &=& \frac{1}{a}[\psi(x)-
  \widetilde{U}^{-1}_\mu (x-a\hat{\mu}) \psi(x-a\hat{\mu})].
\end{eqnarray}
% \begin{equation}
 % M_{xy} = \delta_{xy} - 
  %\kappa \sum_\mu \left[ (1+\gamma_\mu) V_{x,\mu} +
     %    (1-\gamma_\mu)V^T_{x-\mu,\mu}\right].
%\end{equation}
Here, the link variables are the usual ones,
$\widetilde{U}_\mu(x)=U_\mu(x)$, for fermions in the fundamental
representation while for the adjoint representation they are
\begin{equation}
  \widetilde{U}_\mu^{ab}(x)=2 \Tr (T^aU_\mu(x)T^bU_\mu^\dagger(x)),
   \label{eq:adjproject}
\end{equation}
where $T^a$, $a=1,2,3$, are the generators of the fundamental
representation, normalised as $\Tr T^aT^b=\frac12 \delta^{ab}$.
We note that in the adjoint representation
the elements of $\widetilde{U}$-matrices are real and
$\widetilde{U}^{-1} = \widetilde{U}^T$. 

The lattice action (\ref{eq:aktion}) is parametrised with two dimensionless parameters,
$\beta_L = 4/g_{\textrm{bare}}^2$ and $\kappa = 1/[8+2am_{q,0}]$.  The
parameter $\kappa$ is related to the fermion mass.  In the continuum
limit $a^4\sum_x\rightarrow \int d^4x$ as $a\rightarrow 0$, and the
leading order contribution from (\ref{eq:aktion}) yields the continuum
action while the terms of higher order in $a$ will be suppressed;
these terms are generically termed ``lattice artifacts''. Since gauge
invariance forbids any contribution from dimension five operators to
the gauge action, only the fermion action here is subject to lattice
artifacts at ${\mathcal{O}}(a)$. These are 
removed (for on-shell quantities) by considering the improved action 
\begin{eqnarray}
  S_{\rm{impr}} &=& S_0+\delta S_{\textrm{sw}},\\
  \delta S_{\textrm{sw}} &=& a^5\sum_x
  \csw \bar\psi(x)\frac{i}{4}\sigma_{\mu\nu}
  F_{\mu\nu}(x)\psi(x)\label{swterm}
\end{eqnarray} 
and tuning the Sheikholeslami-Wohlert coefficient $\csw$ at each 
$\beta_L$ so that the ${\mathcal{O}}(a)$ effects in on-shell 
quantities cancel; to lowest order in perturbation theory $\csw=1$
\cite{Sheikholeslami:1985ij}. Here
$\sigma_{\mu\nu}=i[\gamma_\mu,\gamma_\nu]/2$ and $F_{\mu\nu}(x)$ is
the ``clover term'', lattice field strength tensor
in the appropriate representation
symmetrized over the 
four $\mu,\nu$-plane plaquettes which include the point $x$.
 
% We note that because the link matrices are real, it would be
% possible to perform lattice simulations with two Dirac flavors of
% staggered fermions without encountering the square root problem (for
% an example, see \cite{Karsch:1998qj}).

Because our aim in future work is to measure the evolution of the
gauge coupling constant using the  Schr\"odinger functional method, we
also need to consider the improvement of the
action at the special Schr\"odinger functional boundary conditions.
Schr\"odinger functional method is also used in this work 
for measuring $\csw$, but for this the boundary improvement
is not necessary.
%In this work we will work exclusively with the Schr\"odinger
%functional formulation: 

We consider a system of 
size $L^3 \times T$, with periodic boundary conditions
to the spatial directions and with 
Dirichlet boundary conditions for the gauge fields 
to the time direction:
\begin{equation}
  U_k(x_0=0)=W(k)
  % \exp(-i\eta\sigma_3 a/L)
  ,\,\,\,U_k(x_0=T)=W^\prime(k),
  % \exp{-i(\pi-\eta)\sigma_3a/L)},
\end{equation}
where $k=1,2,3$; the explicit form of the boundary fields will be
discussed later. 
For the measurement of the coupling constant the 
boundary gauge fields are chosen so that they
lead to a constant background chromoelectric field. Due to the frozen 
boundaries there now exists ${\mathcal{O}}(a)$ contribution
to the gauge part of
the action, and to account for these we consider 
\begin{equation}
  S_{G,{\textrm{impr}}}=\frac{\beta_L}{4}\sum_p
  w(p){\textrm{tr}}(1-U(p)),
\end{equation}
where the weights $w(p)$ are equal to 1 for plaquettes in the bulk,
$w(p)=c_s/2$ for spatial plaquettes at $x_0=0$ and $T$ and $w(p)=c_t$
for time-like plaquettes attached to a boundary plane. The parameters
$c_s$ and $c_t$ are tuned to reduce the ${\mathcal{O}}(a)$ boundary
contributions.%
\footnote{%
  Recall that gauge invariance guarantees that there are no
  ${\mathcal{O}}(a)$ contributions to the gauge action in the bulk,
  and hence the boundary terms controlled by $c_s$ and $c_t$ are the
  only ones which arise to ${\mathcal{O}}(a)$ in the gauge action.}
To leading order in perturbation theory $c_t=c_s=1$. For the electric
background field which we consider the terms proportional to $c_s$ do
not contribute.

The boundary values of the fermion fields are set as
\begin{equation}
\begin{array}{rcl}
  P_+\psi(x_0=0,{\bf{x}}) &=& \rho({\bf{x}}),\,\,\, 
  P_-\psi(x_0=T,{\bf{x}})+\rho^\prime({\bf{x}}),\\
  P_-\psi(x_0=0,{\bf{x}}) &=& P_+\psi(x_0=T,{\bf{x}}) = 0,
\end{array}
\label{eq:boundary_rho}
\end{equation}
with similar definitions on the conjugate fields. 
The projection operators are $P_{\pm}=\frac{1}{2}(1\pm\gamma_0)$. 
The boundary fields $\rho$, $\rho'$ are source fields for
correlation functions, and they are set to zero when generating
configurations in simulations.
In the spatial directions it is customary to introduce
a ``twist'' for the phase of the fermion fields
\cite{Sint:1995rb}:
\begin{equation} \label{eq:boundary_theta}
  \psi(x+L\hat{k})=e^{i\theta_k}\psi(x),\,\,\,
  \bar\psi(x+L\hat{k})=\bar\psi(x)e^{-i\theta_k}.  
\end{equation}
In this work we use $\theta_k = \pi/5$ throughout.
The twist, together with the Dirichlet boundary conditions,
regulates the fermion matrix so that simulations at
zero fermion masses become possible.
%Then the fermion action is given by \eqref{sf} with the above
%definitions of the lattice derivatives.

The improved lattice action is now given by 
\begin{equation}
  S_{\textrm{impr}}=S_{G,{\textrm{impr}}}+S_F+\delta
  S_{\textrm{sw}}+\delta S_{F,\textrm{b}}.
  \label{improvedaction}
\end{equation}
Now the Sheikholeslami-Wohlert term only accounts for the bulk,
\begin{equation}
  \delta S_{\textrm{cw}} = 
  a^5\sum_{x_0=a}^{T-a}\sum_{\bf{x}}c_\textrm{sw} 
  \bar\psi(x)\frac{i}{4}\sigma_{\mu\nu}F_{\mu\nu}(x)\psi(x),
\end{equation}
while the boundary effects are captured by $\delta S_{F,\textrm{b}}$. 
This counterterm has two contributions, controlled
by parameters denoted by $\tilde c_s$ and $\tilde c_t$. The term proportional to $\tilde c_s$ is
\bea
  \delta S_{\tilde{c}_s}&=& 
  a^4 (\tilde{c}_s-1)\sum_{\bf{x}}\bigg[\frac{1}{2}\bar{\psi}(0,{\bf{x}})
  P_-\gamma_k(\nabla^* _k+\nabla_k)P_+\psi(0,{\bf{x}})\nonumber\\
&&+\frac{1}{2}\bar{\psi}(L,{\bf{x}})P_+\gamma_k(\nabla^* _k+\nabla_k)
  P_-\psi(L,{\bf{x}})\bigg]
\eea
and it clearly vanishes if we set fermionic fields to zero on the boundaries.

So, similarly to
the gauge action, only the term proportional to $\tilde c_t$
contributes, and this contribution is given by \be \delta
S_{F,{\textrm{b}}}=a^4\sum_x(\tilde
c_t-1)\frac{1}{a}\bar\psi(x)\psi(x)(\delta(x_0-a)+\delta(x_0-(L-a)).
\ee This can be seen as a correction to the bare mass term at $x_0=a$
and $x_0=L-a$, hence accounted for by the modification \be
m_{q,0}\mapsto m_{q,0}+(\tilde c_t-1)(\delta_{t,a}+\delta_{t,L-a}).
\ee 
It is known that $\tilde c_t=1$ to leading order.\footnote{%
  This
  is so because free Wilson fermions are not subject to
  ${\mathcal{O}}(a)$ artifacts.}

Hence, to obtain ${\mathcal{O}}(a)$ improvement we need to 
determine the parameters $c_t$, $\tilde c_t$ 
and $c_{\textrm{sw}}$ in the action (\ref{improvedaction}). 
The parameters $c_t$ and $\tilde c_t$ are determined perturbatively 
as will be described in the following section. The parameter 
$c_{\textrm{sw}}$ is determined nonperturbatively, and this will 
be determined in section \ref{cswdetermination}.

\section{Perturbative analysis of the boundary improvement}
\label{perturbative}

As explained in the previous section, due to the Dirichlet 
boundary conditions associated with the Schr\"odinger 
functional formalism, we are led to counteract ${\mathcal{O}}(a)$  
lattice artifacts on the boundaries both in the gauge and 
fermion parts of the action. In this section we describe in 
detail the analysis of the rquired counterterms. Although we 
are mostly interested in matter fields in fundamental or 
adjoint representation of SU(2) gauge group, we will present 
the results applicable also for higher representations of SU(3) 
since these are relevant for the current developments in 
the studies of these theories on the lattice. 

In principle there exists four counterterms associated with 
the spatial links in the boundary and with temporal links 
connected to the boundary. Due to the specific form of the 
background field we have chosen, only two of these are 
needed and these are denoted by $c_t$ and $\tilde{c}_t$. 
These boundary coefficients have a perturbative expansion 
of the form
\be
c_x=1+c_x ^{(1)} g_0 ^2+\mathcal{O}(g_0 ^4).
\ee
Our goal is to determine $\tilde{c}_t$ and $c_t$ to one-loop 
order in perturbation theory.

\subsection{Coefficient $\tilde{c}^{(1)}_t$}
\begin{figure}
\begin{center}
\includegraphics[width=0.18\linewidth]{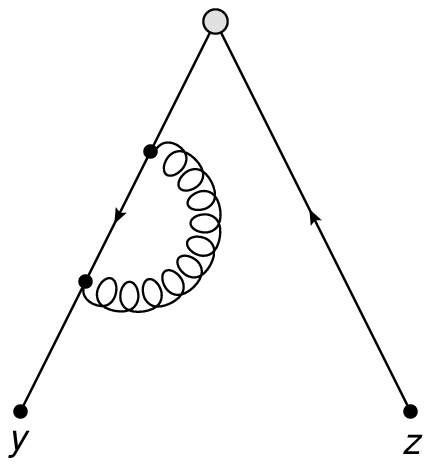}\hspace{1cm}
\includegraphics[width=0.18\linewidth]{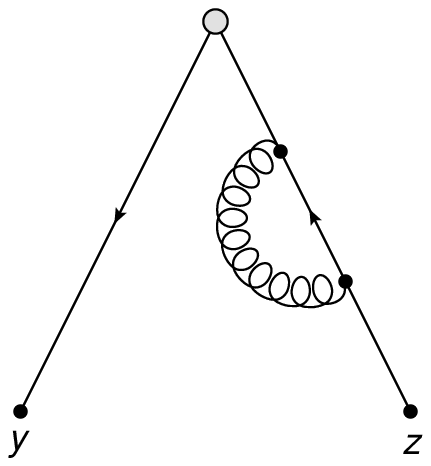}\\
\includegraphics[width=0.18\linewidth]{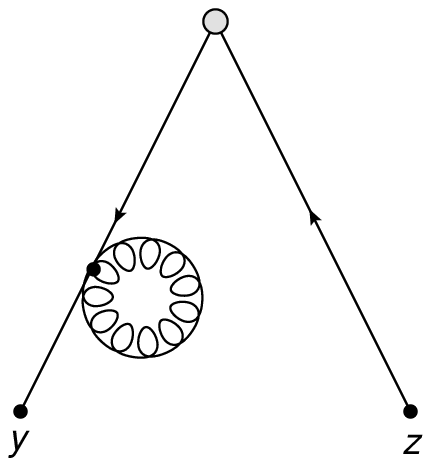}\hspace{1cm}
\includegraphics[width=0.18\linewidth]{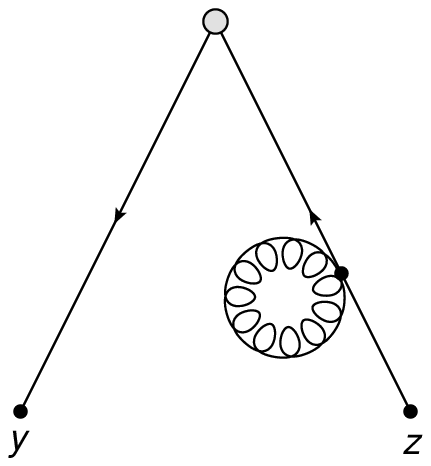}\hspace{1cm}
\includegraphics[width=0.18\linewidth]{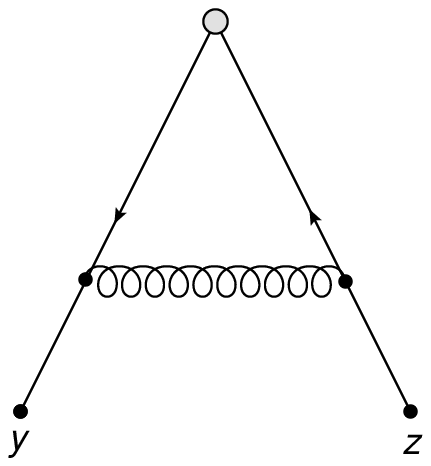}
\end{center}
\caption{Diagrams contributing to the calculation of $\tilde{c}_t^{(1)}$. 
The shaded blob on each diagram indicates the insertion 
of the operator $\Gamma_x=\{{\mathbbm{1}},\gamma_5\}$.}
\label{diagrams}
\end{figure}

We follow the analysis performed in \cite{Luscher:1996vw} 
for the fundamental representation. The result of \cite{Luscher:1996vw} 
is
\be
\tilde{c}_t^{(1)}=-0.0135(1) C_F,
\ee 
and this generalizes to other fermion representations simply 
by replacing the fundamental representation Casimir operator $C_F$ 
with Casimir operator $C_R$ of the representation $R$ under 
consideration. This is so because the relevant correlations functions 
are proportional to the diagrams presented in figure~\ref{diagrams}, 
which all include the color factor $\sum_a (T^a)^2=C_R$. 
Thus it can be shown that also $\tilde{c}_t^{(1)}\propto C_R$.

%\FIGURE{
%}

In the case of fundamental fermions the original result of 
\cite{Luscher:1996vw} is directly applicable with 
$C_F=(N_c^2-1)/(2N_c)=3/4$ for $N_c=2$. 
For the other case we have fermions transforming 
in the adjoint representation of SU(2), for which the 
Casimir invariant is $C_A=2$. The results for different 
gauge groups and fermion representations are 
shown in table~\ref{table:pert_impro}.

\subsection{Coefficient $c^{(1)}_t$}
The coefficient $c^{(1)}_t$ can be split into gauge and fermionic parts 
\be
c_t^{(1)}=c_t^{(1,0)}+c_t^{(1,1)}N_f.
\ee
The contribution $c_t^{(1,0)}$ is entirely due to gauge 
fields and has been evaluated 
%for fundamental representation fermions 
in \cite{Luscher:1992an} for SU(2) and in \cite{Luscher:1993gh} 
for SU(3).  The fermionic contribution $c_t^{(1,1)}$ to $c_t$ 
has been evaluated for fundamental fermions in \cite{Sint:1995ch} 
both for SU(2) and SU(3). We have extended these 
computations for SU(2) and SU(3) gauge theory with higher 
representation fermions and for SU(4) gauge theory 
with fundamental representation fermions. 

The method we have used is the same as the 
one presented in \cite{Sint:1995ch}, with two exceptions. 
First, the boundary fields have to be transformed 
to the desired fermion representation. 
Generally the boundary fields are of the form 
\be 
%\label{eq:abelian1}
U(x,k)|_{(x_0=0)}=\exp(aC_k),\quad U(x,k)|_{(x_0=L)}=\exp(aC'_k),
\ee
where
\be 
%\label{eq:abelian2}
C_k=\frac{i}{L}{\rm{diag}}(\phi_1,\dots,\phi_n),
\quad C'_k=\frac{i}{L}{\rm{diag}}(\phi'_1,\dots,\phi'_n),
\ee   
and $n$ is the dimension of the representation. 
The transformed boundary fields are obtained from the fundamental 
representation counterparts for adjoint representation 
via \eqref{eq:adjproject}. 
After the transformation one simply diagonalizes the 
resulting matrices and ends up with a matrix of the form
\be
{\rm{diag}}\left(\exp[i\phi^A _1],\dots,\exp[i\phi^A _n]\right),
\ee
where $\phi^A _i$ give the adjoint representation boundary fields
\be
C^A _k=\frac{i}{L}{\rm{diag}}(\phi^A _1,\dots,\phi^A _n),
\quad C'_k=\frac{i}{L}{\rm{diag}}(\phi'^A _1,\dots,\phi'^A _n).
\ee

For the symmetric representation the components of 
the boundary fields $C^S _k$ and $C'^S _k$ can be obtained by 
taking all the symmetric combinations of $\phi_i$. For $SU(3)$ 
sextet representation this is
\be\begin{array}{ccc}
\phi^S _1&=&\phi_1+\phi_1,\\
\phi^S _2&=&\phi_1+\phi_2,\\
\phi^S _3&=&\phi_1+\phi_3,\\
\phi^S _4&=&\phi_2+\phi_2,\\
\phi^S _5&=&\phi_2+\phi_3,\\
\phi^S _6&=&\phi_3+\phi_3.
\end{array}
\ee

The other crucial note comes from the normalization in 
the calculation of $c_t ^{(1,1)}$. 
Using the Schr\"odinger functional scheme and taking 
the lattice action with constant background field as an 
effective action $\Gamma_0$, the running coupling is defined via  
\be
\bar{g}^2=\frac{\partial_{\eta}\Gamma_0}{\partial_{\eta}\Gamma}.
\ee
The boundary fields $C_k$ and $C'_k$ are functions of the parameter 
$\eta$ so the running coupling is given by the change of the 
system as the boundary fields are altered. 
The effective action $\Gamma$ is to one loop order in perturbation theory
\be
\Gamma=g_0 ^{-2}\Gamma_0+\Gamma_1+\mathcal{O}(g_0 ^2),
\ee   
so the running coupling can be written, as a function 
of the bare coupling $g_0$, in the form 
\be
\bar{g}^2=g_0 ^2\left(1-g_0 ^2\frac{\partial_{\eta}\Gamma_1}{\partial_{\eta}\Gamma_0}\right)+\mathcal{O}(g_0 ^6).
\ee
On small lattice spacings $a$, the one loop correction 
$\Gamma_1$ diverges. This leads to renormalization of the
lattice coupling, which is given in terms of the bare coupling as
\be
g_{\rm{lat}} ^2=g_0 ^2+z_1 g_0 ^4+\mathcal{O}(g_0 ^6),
\ee
where $z_1=2 b_0 \ln(a\mu)$ and 
\be
b_0=\frac{1}{(4\pi)^2}\left(\frac{11}{3}C_A - \frac{4}{3}T(R)N_F\right)
\ee
is the coefficient in one loop beta function. 
Now we can write the running coupling as a function 
of the renormalized coupling
\be
\bar{g}^2=g_{\rm{lat}} ^2\left[1-g_{\rm{lat}}^2
\bigg(\frac{\partial_{\eta}\Gamma_1}{\partial_{\eta}\Gamma_0}
+z_1\bigg)\right]+\mathcal{O}(g_{\rm{lat}} ^6).
\ee

The one loop correction to the effective action 
$\Gamma_1$ can also be written as 
\be
\Gamma_1=\frac{1}{2}\ln\det\Delta_1-\ln\det\Delta_0-\frac{1}{2}\ln\det\Delta_2,
\ee
where the operators $\Delta_0$ and $\Delta_1$ 
are related to the gauge fixing and pure gauge part 
of the action and the operator $\Delta_2=[(D_{sw}+m_0)\gamma_5]^2$
is related to the fermionic part of the action. The operator $D_{sw}$ 
is the lattice Dirac operator that includes the Sheikholeslami-Wohlert term. 
Now for the calculation of $\Delta_2$ one needs to transform 
the boundary fields to the appropriate representation. 
However in the calculation of $\Gamma_0$ 
one needs to keep the boundary fields in the fundamental 
representation. This is so becouse the pure gauge part 
of $c_t ^{(1)}$ should be independent of the representation 
of the fermions and this can only be achived if the 
boundary fields in $\Gamma_0$ are kept in the 
fundamental representation. Also this produces the 
expected behavior for the series expansion of 
\be
\left(\frac{\partial_{\eta}\Gamma_1}{\partial_{\eta}\Gamma_0}+z_1\right). 
\ee

%\newline
%$SU_{A}(2)$
%\begin{equation}
%\begin{array}{lclclcl}
%\phi_1 & = & -2\eta,&&\phi'_1 &=& 2\eta-2\pi,\\
%\phi_2 & = & 2\eta,&&\phi'_2 &=& -2\eta+2\pi,\\
%\phi_3 & = & 0,&&\phi'_3 &=& 0,
%\end{array}
%\label{SU2_adj}
%\end{equation}
%$SU_{A}(3)$
%\begin{equation}
%\begin{array}{lclclcl}
%\phi_1 & = & 2[\eta(3/2-\nu)-\pi/3],&&\phi'_1 &=& -2[\eta(3/2+\nu)+4\pi/3],\\
%\phi_2 & = & -2[\eta(3/2-\nu)-\pi/3],&&\phi'_2 &=& 2[\eta(3/2+\nu)+4\pi/3],\\
%\phi_3 & = & 2[\eta(3/2+\nu)-2\pi/3],&&\phi'_3 &=& -2[\eta(3/2-\nu)+5\pi/3],\\
%\phi_4 & = & -2[\eta(3/2+\nu)-2\pi/3],&&\phi'_4 &=& 2[\eta(3/2-\nu)+5\pi/3],\\
%\phi_5 & = & 4\eta\nu-2\pi/3,&&\phi'_5 &=& 4\eta\nu-2\pi/3,\\
%\phi_6 & = & -4\eta\nu+2\pi/3,&&\phi'_6 &=& -4\eta\nu+2\pi/3,\\
%\phi_7 & = & 0,&&\phi'_7 &=& 0,\\
%\phi_8 & = & 0,&&\phi'_8 &=& 0,
%\end{array}
%\label{SU3_adj}
%\end{equation}
%$SU_{S}(3)$
%\begin{equation}
%\begin{array}{lclclcl}
%\phi_1 & = & 2[\eta-\pi/3],&&\phi'_1 &=& -2(\eta+\pi),\\
%\phi_2 & = & 2\eta(\nu-1/2),&&\phi'_2 &=& 2[\eta(1/2+\nu)+\pi/3],\\
%\phi_3 & = & -2[\eta(1/2+\nu)-\pi/3],&&\phi'_3 &=& -2[\eta(\nu-1/2)-2\pi/3],\\
%\phi_4 & = & \eta(1/2+\nu)-\pi/3],&&\phi'_4 &=& \eta(\nu-1/2)-2\pi/3],\\
%\phi_5 & = & -\eta(\nu-1/2),&&\phi'_5 &=& -\eta(\nu+1/2),\\
%\phi_6 & = & -\eta+\pi/3,&&\phi'_6 &=& \eta+\pi.
%\end{array}
%\label{SU3_sextet}
%\end{equation}

With these remarks, the numerical calculation is straightforward. 
The results for the nonzero improvement coefficients are tabulated 
in table \ref{table:pert_impro}. The numbers beyond the fundamental 
representation are new, while those for the fundamental 
representation provide a good check on our computations. 
For the application to minimal walking technicolor, 
the relevant numbers are the ones on the second 
line of table~\ref{table:pert_impro}.

Our results are consistent with the generic formula
\be
c_t^{(1,1)} \approx 0.019141(2T(R)),
\label{conjecture}
\ee
where $T(R)$ is the normalization of the representation $R$, 
defined as ${\rm{Tr}}(T^a_R T^b_R)=T(R)\delta^{ab}$. 
For the details of the numerical method used to determine 
coefficient $c_t^{(1,1)}$, we refer to the original literature 
where the method was developed and applied first for 
the pure gauge theory case in \cite{Luscher:1992an}, 
and later for fundamental representation fermions in 
\cite{Luscher:1992an,Sint:1995ch}. 

\TABLE{
%\begin{table}[h!bt]
\centering
\begin{tabular}{|c|c|c|c|c|}
\hline
$N_c$ & rep. & $c_t^{(1,0)}$ & $c_t^{(1,1)}$ &  $\tilde{c}_t^{(1)}$ \\
\hline
2 & ${\bf{2}}$ & $-0.0543(5)$ & $0.0192(2)$ & $-0.0101(3)$ \\
2 & ${\bf{3}}$ & $-0.0543(5)$ & $0.075(1)$ & $-0.0270(2)$ \\
3 & ${\bf{3}}$ & $-0.08900(5)$ & $0.0192(4)$ & $-0.0180(1)$ \\
3 & ${\bf{8}}$ & $-0.08900(5)$ & $0.113(1)$ & $-0.0405(3)$\\
3 & ${\bf{6}}$ & $-0.08900(5)$ & $0.0946(9)$ & $-0.0450(3)$\\
4 & ${\bf{4}}$ &               & $0.0192(5)$ & $-0.0253(2)$\\
\hline
\end{tabular}
\caption{The nonzero improvement coefficients for Schr\"odinger 
functional boundary conditions with electric background field 
for various gauge groups and fermion representations.}
\label{table:pert_impro}
%\end{table}
}

\begin{figure}
\centering
\includegraphics[scale=0.45]{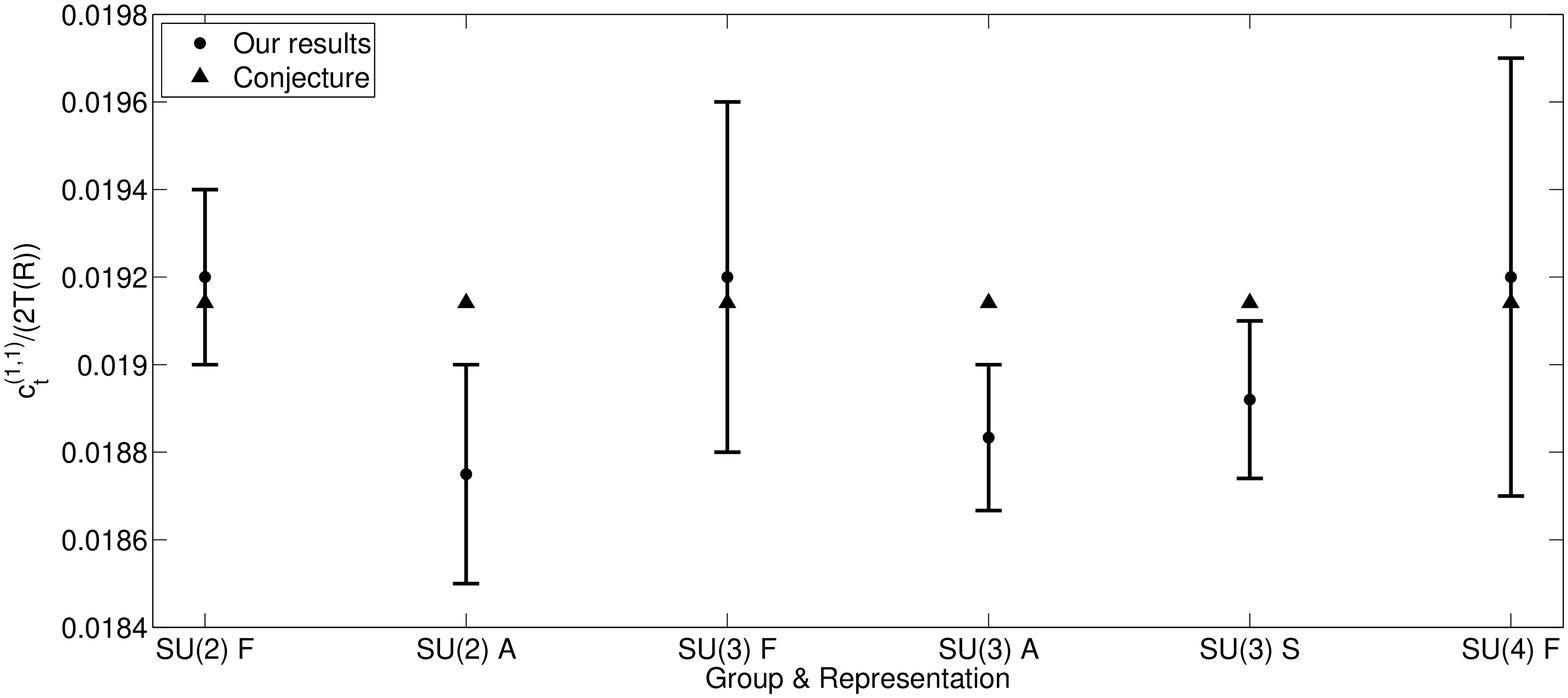}
\caption{Our results of $c_t ^{(1,1)}$ scaled with $2T(R)$ 
compared with conjectured value of $c_t ^{(1,1)}/(2T(R))$.}
\label{Pic}
\end{figure}

We have also plotted our results of $c_t ^{(1,1)}$ scaled with 
$1/(2T(R))$ against \eqref{conjecture} in figure~\ref{Pic}. 
Although we did not achieve the accuracy of the original 
work \cite{Sint:1995ch}, our results are fully compatible for 
fundamental representation fermions. The figure also clearly 
indicates that $c^{(1,1)}_t$ scales with $2T(R)$. 

%\FIGURE{
%}

\section{Non-perturbative tuning}
\label{cswdetermination}

The continuum physics we are interested in corresponds to massless 
fermions, so we need to simulate at zero physical quark mass. With 
Wilson fermions the bare quark mass is additively renormalized, and 
the zero of the physical quark mass corresponds to tuning the bare 
quark mass to a critical value, $m_0=m_{\rm{cr}}$. This tuning is 
done nonperturbatively and allows for determination of the improvement 
coefficient $\csw$ simultaneously.  Here we describe the calculation
of $\csw$ for $N_f=2$ flavors of SU(2) fundamental and adjoint 
representation fermions.

In these simulations the fermion fields have the boundary
conditions given in Eqs.~(\ref{eq:boundary_rho},\ref{eq:boundary_theta}).
For the fundamental representation fermons we fix the gauge field
Dirichlet boundary
conditions at $x_0=0$ and $x_0=T$~\cite{Luscher:1992zx}:
\begin{eqnarray} 
  U_k(x_0=T) &=& \exp(i C'), \,\,\,\, 
  C' = -\frac{\pi}{4} \frac{a\sigma^3}{L} 
  \label{eq:abelian1} \\
  U_k(x_0=0) &=& \exp(i C), \,\,\,\,
  C = -\frac{3\pi}{4} \frac{a\sigma^3}{L},
  \label{eq:abelian2}
\end{eqnarray}
for $k=1,2,3$.  Because the boundary link matrices commute,
we call these boundary conditions Abelian, in contrast to the
non-Abelian (non-commuting) ones defined below.

The physical quark mass is defined via the partial conservation of the axial current (PCAC) relation,
\be
  M(x_0)=\frac{1}{2}\frac{\frac{1}{2}(\partial_0^\ast+\partial_0)f_A(x_0)
    + c_A a\partial_0^\ast\partial_0 f_P(x_0)}
  {f_P(x_0)}  \equiv  r(x_0) +c_A s(x_0),
\label{eq:M}
\ee
where
\bea
A^a_\mu &=& \bar{\psi}(x)\gamma_5\gamma_\mu\frac{1}{2}\sigma^a\psi(x), \\
P^a &=& \bar{\psi}(x)\gamma_5\frac{1}{2}\sigma^a\psi(x),\\
f_A(x_0) &=& -a^6\sum_{\bf{y,z}}\langle A_0^a(x)\bar{\zeta}({\bf{y}})\gamma_5\frac{1}{2}\sigma^a\zeta({\bf{z}})\rangle,\\
f_P(x_0) &=& -a^6\sum_{\bf{y,z}}\langle P^a(x)\bar{\zeta}({\bf{y}})\gamma_5\frac{1}{2}\sigma^a\zeta({\bf{z}})\rangle.
\eea

Another set of correlation functions, $f_A^\prime$ and $f_P^\prime$ is defined via
\bea
f^\prime_A(T-x_0) &=& -a^6\sum_{\bf{y,z}}\langle A_0^a(x)\bar{\zeta}({\bf{y}})\gamma_5\frac{1}{2}\sigma^a\zeta({\bf{z}})\rangle,\\
f^\prime_P(T-x_0) &=& -a^6\sum_{\bf{y,z}}\langle P^a(x)\bar{\zeta}({\bf{y}})\gamma_5\frac{1}{2}\sigma^a\zeta({\bf{z}})\rangle.
\eea

The bare mass is tuned so that $M(T/2)$ vanishes. The $c_{\rm{sw}}$ term is tuned simultaneously using mass measurements at a different point in the bulk looking for variations of the order of the lattice spacing. Defining $M^\prime$ with obvious replacements of primes, it follows that the quantity
\be
\Delta M(x_0)=M(x_0)-M^\prime(x_0)
\label{eq:deltaM}
\ee
vanishes up to corrections of ${\mathcal{O}}(a^2)$ if both $\csw$ and $c_A$ 
have their proper values. In order to recover the correct tree level 
behaviour we fix these quantities $M$ and $\Delta M$ two their tree level 
values, measured by from a cold gauge configuration with $\kappa_c=0.125$. 
This gives a small correction to the relations:
\be \label{eq:DeltaMcondition}
\Delta M(x_0)=M(x_0)-M^\prime(x_0)-\delta = 0, \\
M(x_0) = \delta_M
\ee

However, for the adjoint representation fermions there are
complications which significantly reduce the effectiveness of the
above method.  Using Eq.~\eqref{eq:adjproject} we immediately notice
that the Abelian boundary matrices (\ref{eq:abelian1},\ref{eq:abelian2}) are
transformed into form
\begin{equation}
  \widetilde{U}_{k} =
  \left( \begin{array}{ccc}
      \ldots & \ldots & 0 \\
      \ldots & \ldots & 0 \\
      0 & 0 & 1
    \end{array} \right)
\end{equation}
Thus, there is a component of the adjoint representation color vector 
which simply does not see the background field. This feature is independent
of the color structure chosen for the boundary conditions.
It turns out that regardless
of how the fermion sources or the constant boundary conditions are chosen, 
at long distances the correlation functions behave as if there is no
background field. In other words,
the adjoint fermion correlation functions ``see'' the background 
electric field only at short distances.  This significantly reduces the
effectiviness of the background field method for tuning $\csw$.

This effect can be improved by using boundary conditions
which maximize the difference between the two boundaries. We use the
following asymmetric "non-Abelian" boundary conditions: links at the
upper $x_0=T$ boundary are chosen to be trivial
\begin{equation} 
  \label{non_abelian1}
  U(x_0=T,k) = I
\end{equation}
and at the lower boundary $x_0=0$ we use
\begin{equation} 
  \label{non_abelian2}
  U(x_0=0,k) = \exp(a C_k), \,\,\,\, 
  C_k = \frac{\pi}{2} \frac{\tau^k}{i L}.
\end{equation}
This creates a strong chromomagnetic field at $x_0=0$ boundary.
These boundary conditions do not fully cure the problem, but
nevertheless provide enough leverage so that the
PCAC mass relation can be used to tune $\csw$.

\begin{figure}
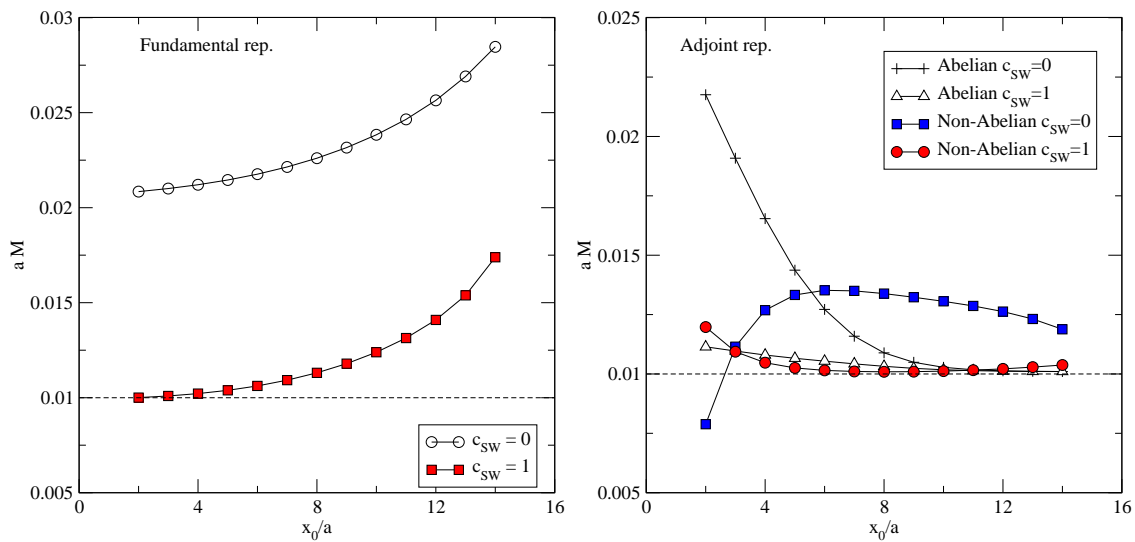

%\FIGURE{
  \centerline{
    \includegraphics[width=.49\linewidth]{m_tree_f.eps}
    \includegraphics[width=.49\linewidth]{m_tree_a.eps}}
  \caption[a]{{\em Left:} Fundamental representation fermion mass
    $aM(x_0)$ measured from the classical gauge field configuration
    satisfying the Abelian boundary conditions
    (\ref{eq:abelian1}, \ref{eq:abelian2}) 
    on a $8^3\times 16$-lattice.  Bare
    mass is $am_0 = 0.01$, which is also $aM$ in the continuum limit.    
    Inclusion of the clover term ($\csw=1$) significantly 
    reduces the cutoff
    effects.  {\em Right:} $aM(x_0)$ for adjoint representation
    fermions and for the Abelian boundary conditions
    (\ref{eq:abelian1}), (\ref{eq:abelian2}), and
    for the non-Abelian boundary conditions
    (\ref{non_abelian1},\ref{non_abelian2}).  
    Here the correlation functions between $\csw=0$
    and $\csw=1$ differ significantly at long distances
    only for the non-Abelian boundary conditions.
  }
  \label{fig:mtree}
%}
\end{figure}

This behaviour can be demonstrated already at the classical 
level: in figure~\ref{fig:mtree} 
we show the PCAC fermion mass \eqref{eq:M}, measured using 
the classical minimum action gauge field configuration which 
satisfies the appropriate boundary conditions.
The bare fermion mass has been set to $am_0 = 0.01$, and,
in the absence of the background field or lattice cutoff effects, 
the PCAC measurement
would yield precisely this value.  However, with finite lattice
spacing the non-trivial
classical background field gives rise
to cutoff effects, which moves the PCAC mass away from $aM = 0.01$.
For the fundamental representation fermions and 
the Abelian boundary conditions (\ref{eq:abelian1}), (\ref{eq:abelian2}).
(left panel in figure~\ref{fig:mtree}),
we can observe that setting $\csw=0$ (non-improved standard
Wilson fermions) the measured mass values are far from
the continuum limit, whereas using $\csw=1$ (the correct
value at the classical level) these effects are strongly
reduced.

For the adjoint representation fermions the behaviour is very different, as shown on the right panel of \fig{fig:mtree}: 
using the Abelian boundary conditions the measured masses $aM(x_0)$ rapidly approach $0.01/a$ as $x_0$ increases, for both $\csw = 0$ or $1$.  This indicates that the correlation function lacks the sensitivity to $\csw$ and cannot be used for tuning it to the correct value.

On the other hand, with the non-Abelian boundary conditions (\ref{non_abelian1},\ref{non_abelian2}) the correlation
function remains sensitive to the value of $\csw$ to 
longer distances.  The sensitivity remains in the
mass asymmetry $\Delta M(x_0)$, \eqref{eq:deltaM},
which can now be used to tune $\csw$.
We note that these boundary conditions are 
useful only for determining $\csw$, not for evaluating the
coupling constant.

In order to remove the dependence on $c_A$, for fundamental fermions, we consider
\be
  M(x_0,y_0)=r(x_0)-s(x_0)\frac{r(y_0)-r^\prime(y_0)}{s(y_0)-s^\prime(y_0)},
\ee
which coincides with $M(x_0)$ up to ${\mathcal{O}}(a^2)$ corrections and is 
independent of $c_A$.
With adjoint fermions this quantity suffers from large 
statistical fluctuations and is not useful. Instead we simply consider 
the quantity $M(x_0)$ and fix $c_A$ to its perturbative value 
\cite{Luscher:1996vw} 
\begin{equation}
  c_A = -0.00567(1) C_R g^2 + \mathcal{O}(g^4).
\end{equation}
We then measure $c_A$ separately to confirm the validity of our choice.

In order to evaluate $\csw$ we used the following routine: we
choose lattice volume $L^3\times T = 8^3\times 16$ for both
fundamental and adjoint representation fermions, and a set of values
of the lattice coupling $\beta$. For fundamental fermions we measure
$M = M(T/2,T/4)$ and $\Delta M = \Delta M(3T/4,T/4)$. For adjoint 
fermions we measure $M = M(T/2)$ and $\Delta M = \Delta M(3T/4)$ fixing 
$c_A$ to its perturbative value.
\begin{enumerate}
\item For a given $\beta$, we choose initial $\csw$ (typically
  extrapolating from results obtained with previous values of
  $\beta$).

\item We choose a couple of values for $\kappa$,
  and determine by interpolation the critical value
  $\kappa_c(\beta,\csw)$ where the fermion mass $M$
  is equal to the tree level value.
  
\item Once we have an estimate of the critical $\kappa$, we choose a
  new value for $\csw$ and repeat the search of $\kappa_c$.

\item At the same time, we measure $\Delta M(\csw)$.  Now we can
  linearly interpolate/extrapolate in $\csw$ so that $\Delta M$
  vanishes, obtaining the desired value of $\csw$.
  Using simulations at this final $\csw$ we can relocate the critical
  $\kappa$, if desired, and verify the results of the interpolation.
\end{enumerate}

The above tuning is done at small $L/a$, and the results are applied for 
all lattice sizes since the $L/a$ dependence is expected to be weak. 
Furthermore, we only consider a range of $\beta$ and fit the critical 
values to an interpolating function to obtain $m^c(\beta_L)$ and 
$\csw^c(\beta_L)$.

\subsection{Measurement of $\csw$}

\begin{figure}
%\FIGURE{
  \centerline{\includegraphics[width=.6\linewidth]{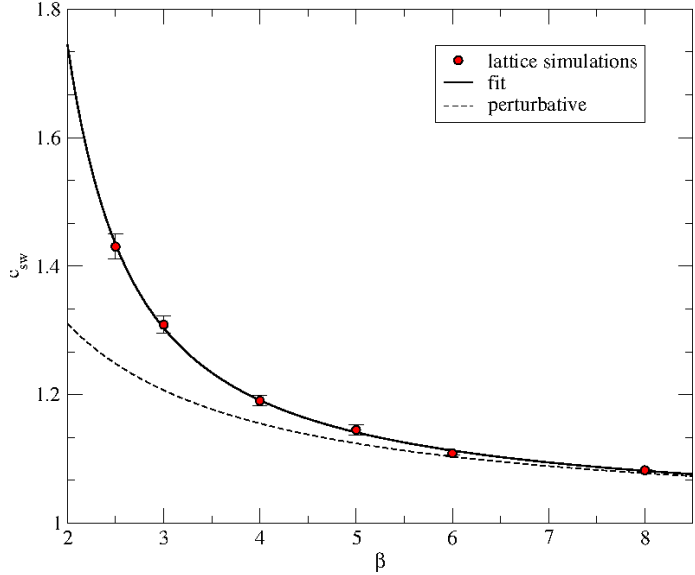}}
  \caption[a]{$\csw$ for two flavors of fundamental 
    representation fermions.  The solid line is the
    interpolating fit, Eq.~(\ref{fundfit}), and
    the dashed line is the 1-loop perturbative
    value}
  \label{fig:fund}
%}
\end{figure}

\begin{figure}
%\FIGURE{
  \centerline{\includegraphics[width=.65\linewidth]{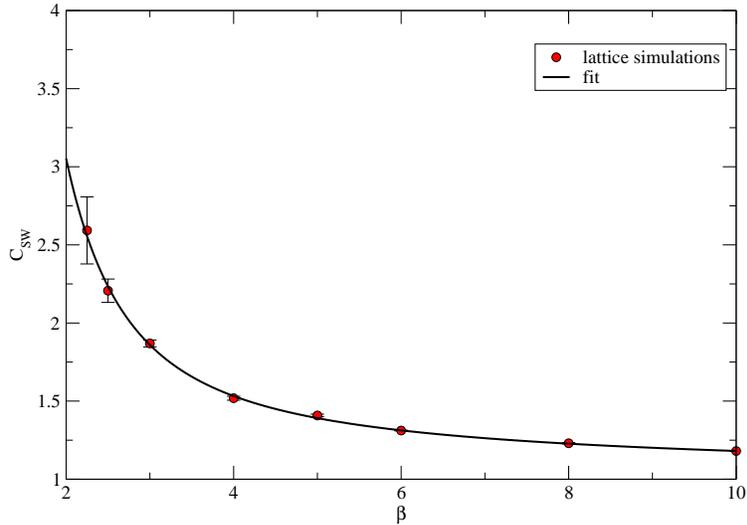}}
  \caption[b]{$\csw$ for two flavors of adjoint 
    representation fermions, with
    the interpolating fit, Eq.~(\ref{adjfit}).}
  \label{fig:adj}
%}
\end{figure}

\begin{table}
%\TABLE{ 
\centering
\begin{tabular*}{0.75\textwidth}{@{\extracolsep{\fill}} | c | c | c | c | c | }
     \hline
	 $\beta$ & $\csw$ & $\kappa$ & $aM$ & $a \Delta M$  \\ \hline \hline
	 $10$ & $1.16$ & $0.1302552$ & $0.00020(7)$ & $-0.0004(1)$  \\
	 $10$ & $1.17208$ & $0.1301818$ & $0.00114(7)$ & $-0.0002(1)$  \\
	 $10$ & $1.1774$ & $0.1301818$ & $0.00050(7)$ & $0.0001(1)$  \\
	 $10$ & $1.17915$ & $0.13017157$ & $0.00037(10)$ & $0.0002(2)$  \\ \hline
	 $8$ & $1.2$ & $0.13171$ & $-0.00156(8)$ & $-0.0004(2)$  \\
	 $8$ & $1.225$ & $0.13154$ & $0.00031(8)$ & $-0.0001(2)$  \\
	 $8$ & $1.227$ & $0.1315265$ & $0.00035(8)$ & $0.0000(2)$  \\
	 $8$ & $1.23$ & $0.1315265$ & $-0.00018(9)$ & $0.0000(2)$  \\
	 $8$ & $1.25$ & $0.1315265$ & $0.00003(8)$ & $0.0003(2)$  \\ \hline
	 $6$ & $1.28$ & $0.1340604$ & $-0.00054(7)$ & $-0.0007(1)$  \\
	 $6$ & $1.3$ & $0.133903$ & $0.00034(8)$ & $-0.0003(1)$  \\
	 $6$ & $1.3135$ & $0.1338131$ & $0.00055(8)$ & $-0.0001(1)$  \\
	 $6$ & $1.3143$ & $0.1338131$ & $0.0002510)$ & $0.0001(1)$  \\
	 $6$ & $1.33$ & $0.1338131$ & $-0.00280(8)$ & $0.0005(1)$  \\ \hline
	 $5$ & $1.3$ & $0.1363278$ & $0.0006(1)$ & $-0.0015(3)$  \\
	 $5$ & $1.4$ & $0.1356033$ & $0.0007(1)$ & $-0.0003(3)$  \\
	 $5$ & $1.4058$ & $0.136$ & $-0.0130(2)$ & $0.0000(3)$  \\
	 $5$ & $1.5$ & $0.1348774$ & $0.0007(1)$ & $0.0014(3)$  \\ \hline
	 $4$ & $1.45$ & $0.1391039$ & $0.0012(2)$ & $-0.0008(3)$  \\
	 $4$ & $1.522$ & $0.1385882$ & $-0.0024(2)$ & $0.0001(2)$  \\
	 $4$ & $1.6$ & $0.1378078$ & $0.0004(2)$ & $0.0008(2)$  \\ \hline
	 $3$ & $1.6$ & $0.145311$ & $0.0002(2)$ & $-0.0022(4)$  \\
	 $3$ & $1.75$ & $0.1435289$ & $0.0038(2)$ & $-0.0005(3)$  \\
	 $3$ & $1.834$ & $0.1426551$ & $0.0018(2)$ & $-0.0006(4)$  \\
	 $3$ & $1.9$ & $0.1419574$ & $0.0009(3)$ & $0.0002(4)$  \\
	 $3$ & $2.1$ & $0.1400727$ & $0.0082(2)$ & $0.0016(3)$  \\ \hline
	 $2.5$ & $1.5$ & $0.1540744$ & $0.0021(4)$ & $-0.023(5)$  \\
	 $2.5$ & $2$ & $0.147733$ & $-0.0036(3)$ & $-0.0005(4)$  \\
	 $2.5$ & $2.5$ & $0.141683$ & $0.0015(2)$ & $0.0005(4)$  \\
	 $2.5$ & $2.7$ & $0.139561$ & $-0.0025(2)$ & $0.0027(9)$  \\ \hline
	 $2.25$ & $1.5$ & $0.1590893$ & $0.0306(3)$ & $-0.0019(6)$  \\
	 $2.25$ & $2.3$ & $0.147733$ & $-0.0004(3)$ & $-0.0004(5)$  \\
	 $2.25$ & $2.5$ & $0.141683$ & $0.0033(3)$ & $-0.0002(4)$  \\
	 \hline
\end{tabular*}
\caption{Results for the quark mass $M$ and $\Delta M$
with two fermions in the adjoint representation}
\label{fig:csw_tulos_taulukko} 
\end{table}

\begin{table}
%\TABLE{ 
\centering
\begin{tabular*}{0.75\textwidth}{@{\extracolsep{\fill}} | c | c || c | c | }
     \hline
	 $\beta$ & $\csw$ & $\beta$ & $\csw$  \\ \hline
	 $8$ & $1.082(2)$ & $4$ & $1.190(8)$  \\
	 $6$  & $1.109(3)$ & $3$ & $1.309(13)$  \\
	 $5$  & $1.145(8)$ & $2.5$ & $1.430(19)$  \\
	 \hline 
\end{tabular*}
\caption{Results for $\csw$ with 
two flavors of fermions in the fundamental representation}
%}
\label{fig:fund_csw_taulukko}
\end{table}

\begin{table}
%\TABLE{ 
\centering
\begin{tabular*}{0.75\textwidth}{@{\extracolsep{\fill}} | c | c || c | c | }
     \hline
	 $\beta$ & $\csw$ & $\beta$ & $\csw$  \\ \hline
	 $10$ & $1.159(3)$ & $4$ & $1.476(17)$  \\
	 $8$  & $1.197(8)$ & $3$ & $1.805(23)$  \\
	 $6$  & $1.291(3)$ & $2.5$ & $2.059(74)$  \\
	 $5$  & $1.376(9)$ & $2.25$ & $2.593(215)$  \\
	 \hline 
\end{tabular*}
\caption{Results for $\csw$ with 
two flavors of fermions in the adjoint representation}
%}
\label{fig:adj_csw_taulukko}
\end{table}

In figures~\ref{fig:fund} and \ref{fig:adj} we show our results for the 
clover coefficient $\csw$ for both fundamental and adjoint representations. 
The values of $\beta$ used are $\beta = 2.5, 3, 4, 5, 6, 8$, 
and also $\beta=2.25$ and $10$ for the adjoint representation. To clarify 
the tuning method we provide the measurements of $M$ and $\Delta M$ with 
adjoint fermions in table \ref{fig:csw_tulos_taulukko}. In tables 
\ref{fig:fund_csw_taulukko} and \ref{fig:adj_csw_taulukko} we give our 
results for $\csw$ for fundamental and adjoint fermions respectively.

Finally, the measured values for $\csw$ can be fitted 
with a rational interpolating expression, which can used 
in simulations for this range of $\beta$-values.  
For fundamental representation fermions we use the
perturbative 1-loop result $\csw = 1 + 0.1551(1) g^2 + O(g^4)$
\cite{Luscher:1996vw} to constrain the fit:
\begin{equation}
\csw= 
\frac{1-0.090254g^2-0.038846g^4+0.028054g^6}{1-(0.1551+0.090254)g^2}.
\label{fundfit}
\end{equation}
For the adjoint representation the perturbative result is not known,
and we obtain the fit result
\begin{equation}
\csw = 
\frac {1+0.032653 g^2 -0.002844 g^4}{1-0.314153 g^2}.
\label{adjfit}
\end{equation}
In both cases the interpolating fits are valid for $\beta \gsim 2.5$.
For the adjoint fermions it is difficult to reach smaller $\beta$-values
because $\csw$ grows rapidly, and while we were able to reach 
$\beta=2.25$ the errors were too large to constrain the fit (\ref{adjfit})
further.

\subsection{Non-Perturbative measurement of $c_A$}

\begin{figure}
%\FIGURE{
  \centerline{\includegraphics[width=.65\linewidth]{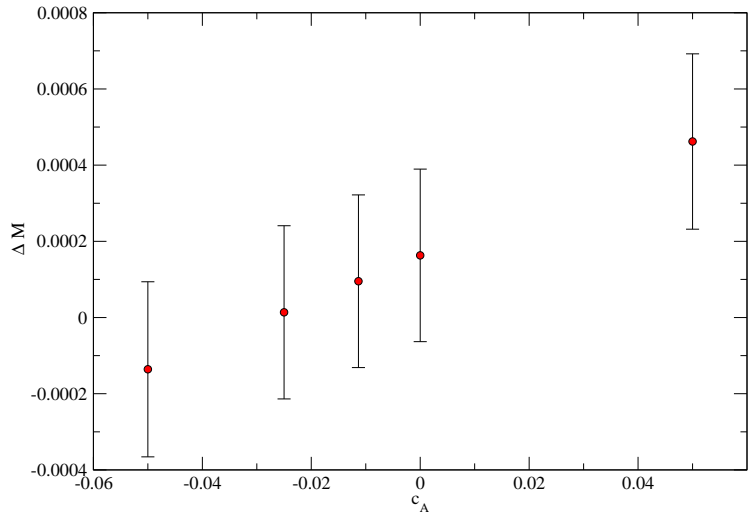}}
  \caption[a]{The dependence of $\Delta M(3T/4)$ of $c_A$. The measurement 
was done with $\beta=4$ and $\csw=1.522$. }
  \label{fig:cAdependence}
%}
\end{figure}

When measuring $\csw$ for adjoint fermions we chose to keep the coefficient 
$c_A$ at the perturbative value. In figure~\ref{fig:cAdependence} we show 
how the choice of $c_A$ affect $\Delta M(3T/4)$ with certain choice of 
parameters. Typically $c_A$ is between $-0.005$ and $-0.01$ at the range of 
$\beta$ we explored. We see that even differences of this order have small 
effect to $\Delta M$.

To verify the accuracy of our choice we have also estimated a non-perturbative 
value for $c_A$. For this we have used the same improvement condition 
as in \cite{Luscher:1996ug}. We do simulations with two different values 
of the fermion phase $\theta$ in the boundary conditions \ref{eq:boundary_theta}, 
using the measured values of $\csw$ and $\kappa_c$ above. Without any 
discretisation errors the difference in the measured masses should be 
equal to the tree level value. Requiring that this condition is met, 
we can find an estimate of $c_A$.

From two simulations with $\theta=0$ and $\theta=\pi/2$ we calculated 
the discretisation effect
\begin{align}\label{c_A_dM}
\Delta M(c_A)' = M\left (x_0=8;\theta=0,c_A\right ) - M\left (x_0=8;\theta=\pi/2,c_A\right ) - \delta,
\end{align}
where $\delta$ is the tree level value of the difference. It is similar 
to the tree level correction in equation \ref{eq:DeltaMcondition} and is relatively 
small. These simulations were done using a trivial boundary condition, 
where all the boundary matrices were set to unity. Depending on the 
lattice coupling between 2000 and 35000 trajectories were performed for 
each value of $\theta$.

As the quark mass, and therefore $\Delta M'$ is simply linearly dependent 
on $c_A$, we can measure $\Delta M'$ for two of values of 
$c_A$ to find the correct value where $\Delta M' = 0$.

\TABLE{ \centering
\begin{tabular*}{0.75\textwidth}{@{\extracolsep{\fill}} | c | c || c | c | }
     \hline
     $\beta$ & $c_A$        &$\beta$ & $c_A$  \\ 
     \hline
     $10$ & $-0.0043(5)$    & $4$    & $-0.0092(12)$    \\
     $8$  & $-0.0056(4)$    & $3$    & $-0.0114(13)$   \\
     $6$  & $-0.0053(6)$    & $2.5$  & $-0.0244(23)$    \\
     $5$  & $-0.0087(5)$    & &   \\
	 \hline 
\end{tabular*}
\caption{Results for $c_A$}
\label{fig:c_A_taulukko}
}

\begin{figure}
%\FIGURE{
  \centerline{\includegraphics[width=.65\linewidth]{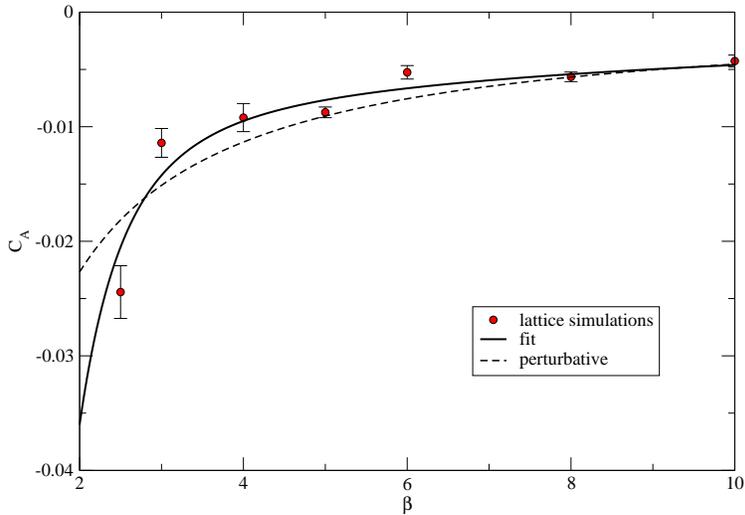}}
  \caption[a]{$c_A$ for two flavors of adjoint fermions. The solid line is the
    interpolating fit, Eq.~(\ref{fundfit}), and
    the dashed line is the 1-loop perturbative
    value}
  \label{fig:cA}
%}
\end{figure}

The results for $c_A$ are given in table \ref{fig:c_A_taulukko} and 
depicted in figure~\ref{fig:cA}. We see that in the region 
where we have measured $c_{sw}$ and $c_A$, it is justified to use the 
perturbative value for $c_A$.

\section{Conclusions and outlook}
\label{outlook}

We have calculated $\cal{O}$-improvement of SU(2) gauge theory with two Wilson fermions
in the fundamental or adjoint representation.  The main results are the
non-perturbative evaluation of the Sheikholeslami-Wohlert clover coefficient 
$\csw$ and the perturbative calculation of
the boundary improvement terms needed for full improvement in the 
Schr\"odinger functional formalism.
The result for $\csw$ is generally applicable to lattice simulations
of these theories.  We also verified that the axial current
improvement coefficient $c_A$ is well described by the 
1-loop perturbative formula in the range of lattice spacings studied.
In addition to the perturbative results on SU(2) gauge theory and adjoint 
fermions, we obtained results also for SU(3) and adjoint or sextet fermions which 
will be useful also for other groups studying these theories.

The main application for the improved action is more accurate 
lattice Monte Carlo analyses of the candidate theory
for minimal walking technicolor,  SU(2) gauge theory with
two adjoint representation fermions.  The boundary improvement
terms permit improved measurement of the evolution of the
coupling constant with the Schr\"odinger functional scheme.
Indeed, in earlier unimproved analyses 
\cite{Hietanen:2009az,DelDebbio:2009fd}
significant cutoff effects were observed at coarse lattices.
The measurement of the coupling with the improved action
is left for future work.

\acknowledgments 
We thank R.~Sommer and S.~Sint for discussions and comments.
This work is supported by the Academy of Finland
grant 114371.  The simulations were performed at the Finnish IT
Center for Science (CSC), Espoo, Finland, and at 
EPCC, University of Edinburgh. 
Parts of the simulation program have been derived from
the MILC lattice simulation program \cite{MILC}.

\end{document}